\documentclass[twocolumn,superscriptaddress,longbibliography]{revtex4-1}%%,twocolumn
\usepackage{graphicx}
\usepackage{times}
\usepackage{amsmath}
\usepackage{amssymb}
\usepackage{units}
\usepackage{stmaryrd} %% "// sign" = \sslash
\usepackage[compact]{titlesec}
\titlespacing{\section}{0pt}{*0}{*0}
\titlespacing{\subsection}{0pt}{*0}{*0}
\titlespacing{\subsubsection}{0pt}{*0}{*0}

\begin{document}
\preprint{0}

\title{Trigger of the ubiquitous surface band bending in 3D topological insulators}

\author{E. Frantzeskakis}
\thanks{emmanouil.frantzeskakis@csnsm.in2p3.fr} 
\address{Van der Waals - Zeeman Institute, Institute of Physics (IoP), University of Amsterdam, Science Park 904, 1098 XH, Amsterdam, the Netherlands}
\address{CSNSM, Universit\'e Paris-Sud, CNRS/IN2P3, Universit\'e Paris-Saclay, 91405 Orsay cedex, France}

\author{S. V. Ramankutty}
\address{Van der Waals - Zeeman Institute, Institute of Physics (IoP), University of Amsterdam, Science Park 904, 1098 XH, Amsterdam, the Netherlands}

\author{N. de Jong}
\address{Van der Waals - Zeeman Institute, Institute of Physics (IoP), University of Amsterdam, Science Park 904, 1098 XH, Amsterdam, the Netherlands}

\author{Y. K. Huang}
\address{Van der Waals - Zeeman Institute, Institute of Physics (IoP), University of Amsterdam, Science Park 904, 1098 XH, Amsterdam, the Netherlands}

\author{Y. Pan}
\address{Van der Waals - Zeeman Institute, Institute of Physics (IoP), University of Amsterdam, Science Park 904, 1098 XH, Amsterdam, the Netherlands}

\author{A. Tytarenko}
\address{Van der Waals - Zeeman Institute, Institute of Physics (IoP), University of Amsterdam, Science Park 904, 1098 XH, Amsterdam, the Netherlands}

\author{M. Radovic}
\address{Swiss Light Source, Paul Scherrer Institut, CH-5232 Villigen, Switzerland}

\author{N. C. Plumb}
\address{Swiss Light Source, Paul Scherrer Institut, CH-5232 Villigen, Switzerland}

\author{M. Shi}
\address{Swiss Light Source, Paul Scherrer Institut, CH-5232 Villigen, Switzerland}

\author{A. Varykhalov}
\address{Helmholtz-Zentrum Berlin f\"{u}r Materialien und Energie, Elektronenspeicherring BESSY II, Albert-Einstein-Strasse 15, 12489 Berlin, Germany}

\author{A. de Visser}
\address{Van der Waals - Zeeman Institute, Institute of Physics (IoP), University of Amsterdam, Science Park 904, 1098 XH, Amsterdam, the Netherlands}

\author{E. van Heumen}
\address{Van der Waals - Zeeman Institute, Institute of Physics (IoP), University of Amsterdam, Science Park 904, 1098 XH, Amsterdam, the Netherlands}

\author{M. S. Golden}
\thanks{m.s.golden@uva.nl}
\address{Van der Waals - Zeeman Institute, Institute of Physics (IoP), University of Amsterdam, Science Park 904, 1098 XH, Amsterdam, the Netherlands}

%\date{\today}

\begin{abstract}
The main scientific activity in the field of topological insulators (TIs) consists of determining their electronic structure by means of magneto-transport and electron spectroscopy with a view to devices based on topological transport. There is however a caveat in this approach. There are systematic experimental discrepancies on the electronic structure of the most pristine surfaces of TI single crystals as determined by Shubnikov de Haas (SdH) oscillations and by Angle Resolved PhotoElectron Spectroscopy (ARPES). We identify intense ultraviolet illumination -that is inherent to an ARPES experiment- as the source for these experimental differences. We explicitly show that illumination is the key parameter, or in other words the trigger, for energetic shifts of electronic bands near the surface of a TI crystal. This finding revisits the common belief that surface decoration is the principal cause of surface band bending and explains why band bending is not a prime issue in the illumination-free magneto-transport studies. Our study further clarifies the role of illumination on the electronic band structure of TIs by revealing its dual effect: downward band bending on very small timescales followed by band flattening at large timescales. Our results therefore allow us to present and predict the complete evolution of the band structure of TIs in a typical ARPES experiment. By virtue of our findings, we pinpoint two alternatives of how to approach flat band conditions by means of photon-based techniques and we suggest a microscopic mechanism that can explain the underlying phenomena.
\end{abstract}

\maketitle

\section{Introduction}

New states of matter in which broken symmetry, topology and spin play a central role can lead to emerging applications in the fields of spintronics and quantum computation. Prime examples of relevant materials platforms in which to realize such novel fermionic degrees of freedom include topological insulators \cite{Hasan2010, Fu2007, Xia2009, Chen2009, Zhang2009, Chen2010, Bianchi2010, King2011}, topological crystalline insulators \cite{Fu2011, Tanaka2012, Dziawa2012, Slager2013} and the recently discovered Weyl semimetals \cite{Liu2014, Neupane2014, Xu2015}. Among these compounds, Bi-based 3D topological insulators (TIs) have attracted enormous scientific interest due to their simple electronic band structure at the Fermi level comprising a single, surface-localized state with linear energy-momentum dispersion \cite{Xia2009, Chen2009, Zhang2009, Hsieh2009, Chen2010}. This so-called Dirac cone harbors spin-polarized electronic states which cross at a special, symmetry-protected spin-degenerate point, known as the Dirac point.

The energy of the Dirac point ($E_{\textmd{D}}$) and its tunability are crucial factors as regimes of topological or trivial transport can be obtained, depending on whether -or not- other electronic states are crossing the Fermi level. Moreover, the possibility of tuning $E_{\textmd{D}}$ in the bulk energy gap, in combination with the large Fermi wavelength for the Dirac fermions, paves the way towards topological superconductors and new fundamental excitations known as Majorana zero modes \cite{Stern2013, Albrecht2016, Mourik2012}. Tunability of the Dirac energy with respect to the Fermi energy ($E_{\textmd{F}}$) has been previously demonstrated in Bi-based TIs via controlled surface decoration \cite{Bianchi2011, King2011, Benia2011, Zhu2011, Bahramy2012, Chen2012, Bianchi2012-2, Valla2012, Zhou2012, Scholz2012, deJong2015}, changes in bulk stoichiometry \cite{Jia2011, Ren2010, Arakane2012, Neupane2012, Miyamoto2012, Zhou2012, Pan2014, Pan2016} and exposure to intense UV illumination \cite{Kordyuk2011, Jiang2012, Frantzeskakis2015}. 

Most of the experimental results on the electronic band structure of TIs come from the direct view of the energy-momentum dispersion provided by Angle Resolved PhotoElectron Spectroscopy (ARPES) \cite{Xia2009, Hsieh2009, Bianchi2010, King2011, Bahramy2012, Benia2011, Zhu2011, Chen2012, Valla2012, Wray2011, Kordyuk2011, Petrushevsky2012, Analytis2010, Frantzeskakis2015, Scholz2012, Bianchi2012, Zhou2012, Hsieh2009-2, Chen2009, Noh2008, Scholz2013, Arakane2012, Neupane2012, Miyamoto2012, Cao2014, Pan2014, Bao2012} or from an extraction of the Fermi surface size using the frequency of quantum oscillations (QO) observed in magneto-transport experiments \cite{Petrushevsky2012, Yan2014, Taskin2012, Taskin2012-2, Kim2014, He2012, Analytis2010-2, Qu2012, Lukyanova2014, Qu2010, Ren2010, Xiong2012, Xiong2012-2, Cao2014, Taskin2011, Hsiung2013, Pan2016, Bao2012}. Although both experimental techniques are very accurate in determining the details of the electronic band structure of TIs (for QOs only for $E\textmd{=}E_{\textmd{F}}$), there is a systematic mismatch in the acquired results. Namely, ARPES results tend to place the Dirac point, $E_{\textmd{D}}$, further below the Fermi level than QO data from the same compounds. In the case of the surface states of 3D TIs, there cannot be a surface vs. bulk discrepancy, as - when executed well on sufficiently bulk-insulating samples - the QO data are also from the surface of the sample.
The upshot is that the ARPES experiments appear more sensitive to band bending effects induced by unwanted surface decoration from residual gas atoms. This is something of a mystery, since ARPES measurements take place in ultrahigh vacuum (UHV) conditions, i.e. under a pressure that is several orders of magnitude lower than that typically found in magneto-transport experiments. Most importantly, the mismatch in the results between spectroscopy and QOs raises serious questions as to whether the results and conclusions obtained by the dominating spectroscopic probe of topological surface states, namely ARPES, can be safely ported to the world of transport-based devices based on topological insulators. 

\begin{figure*}
  \centering
  \includegraphics[width = 16.5 cm]{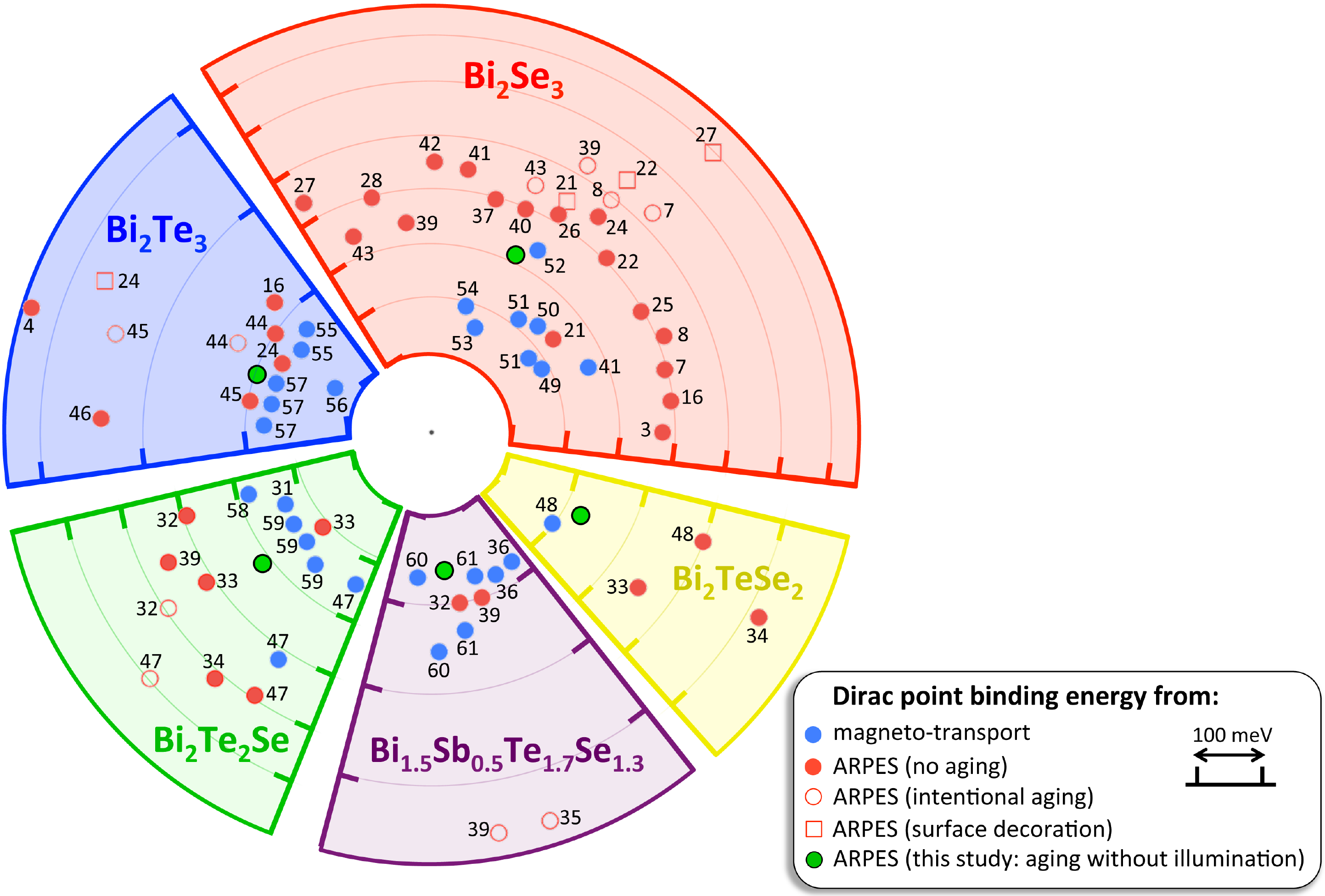}
  \caption{\textbf{Discrepancies between the experimentally determined Dirac point energies for TI surface states between ARPES and quantum oscillations.} The graphic shows the binding energy of the Dirac point ($E_{\textmd{D}}$) in five different 3D topological insulator materials as extracted from published transport (blue symbols) and ARPES (red symbols) studies. The border of the inner circle corresponds to zero binding energy of the Dirac point (i.e. $E_{\textmd{D}}\textmd{=}E_{\textmd{F}}$). The binding energy increases radially with each tick denoting an increase of 100 meV. Notice the different axes scale for each compound. Filled circles denote data from samples which have not been intentionally aged in UHV, and measurements have been typically performed within an hour of sample cleavage. Empty squares and circles denote that air exposure and intentional aging under UHV conditions have been carried out, respectively. Green circles show the value of $E_{\textmd{D}}$ extracted from ARPES data presented in this study: these surfaces have been intentionally exposed to UHV residual gases for 3 hours but their total exposure to external illumination was of only a single second. The green data points are taken from the ARPES data presented in Figs. 2 and 3 for each compound, whereby we note a slight variation in the composition of the Bi$_{2\textmd{-x}}$Sb$_{\textmd{x}}$Te$_{3\textmd{-y}}$Se$_{\textmd{y}}$ used (i.e. $x\textmd{=}0.54$ instead of $0.5$), and that the green ARPES data point in the Bi$_{2}$TeSe$_{2}$ panel is from the related system BiSbTeSe$_{2}$. The number next to each red and blue marker is the literature reference from which $E_{\textmd{D}}$ is extracted.}
\label{fig1}
\end{figure*}

In this paper, we discuss the discrepancies between the two experimental techniques and we present new ARPES data that reveal the drivers behind them. This allows the proposition of an ARPES protocol that enables trustworthy extraction of $E_{\textmd{D}}$ values that are relevant to those obtained in transport studies. In order to reach this goal, we track the complete evolution of the band structure in several different TI compounds during a typical ARPES experiment. In contrast to the common belief that surface decoration is solely responsible for the initial downward band bending, we reveal that the real trigger is illumination. These results point to a simple mechanism that can explain the complex evolution of the electronic band structure of TIs at different timescales during exposure to intense EUV illumination. The proposed mechanism boils down to the interplay of fundamental microscopic processes such as molecular adsorption, photo-dissociation (photolysis), photo-ionization, photo-stimulated desorption and surface photovoltage.\\

\section{Results}

We start by discussing the systematic discrepancy between results on comparable TI single crystals obtained by means of ARPES and transport experiments. The radial scatter plot of Fig. 1 compares the binding energy of the Dirac point obtained either by ARPES experiments (red circles) or by Shubnikov de Haas (SdH) oscillations in magneto-transport (blue circles). The value of $E_{\textmd{F}}\textmd{-}E_{\textmd{D}}$ (i.e. the Dirac point binding energy) increases radially: the border of the inner circle corresponds to zero binding energy of the Dirac point (i.e. $E_{\textmd{D}}\textmd{=}E_{\textmd{F}}$) and each tick denotes an increase of 100 meV. Each data point in the figure corresponds to a different experimental study in the literature, showing the work of many groups, including our own, and results are shown for five different TI compounds. A general conclusion can be readily made. ARPES shows a systematically higher binding energy for the Dirac point than magneto-transport experiments. We note that several ARPES studies \cite{Noh2008, Bianchi2010, Bianchi2011, King2011, Benia2011, Zhu2011, Bahramy2012, Chen2012, Bianchi2012, Arakane2012, Valla2012, Scholz2012, deJong2015, Frantzeskakis2015} have observed energy shifts to higher binding energies because of surface band bending on intentional and unintentional ($\textmd{=}$ `aging') surface decoration. In order to maintain a fair comparison with magneto-transport, the filled red circles in Fig. 1 correspond to surfaces that have been neither decorated nor aged in UHV. Such data points have been acquired in a time frame between a few minutes and 2 hours after cleavage. Empty markers show the value of $E_{\textmd{D}}$ -by means of ARPES- on exposure to air (empty squares) or on increasing exposure to the residual UHV gases (empty circles). Such surface decoration might be an even more important issue in magneto-transport experiments, as such experiments do not take place in a UHV environment and generally do not involve in-situ cleavage of the single crystalline sample. However, the magneto-transport data seems relatively insensitive to surface decoration as the binding energies of the Dirac point are smaller than even the most pristine surfaces studied by ARPES.

Fig. 1 makes it clear that surface decoration alone cannot be the key to the observed differences between ARPES and QO experiments, and thus the conclusion drawn earlier - that the $E_{\textmd{D}}$ values obtained by SdH oscillations cannot be systematically reproduced by ARPES even in the most pristine surfaces - is still valid. In the following, we will explain where the difference in the experimentally determined $E_{\textmd{D}}$ comes from between the two techniques, and we will discuss whether we can approach the SdH values by means of ARPES.
\begin{figure}[!b]
  \centering
  \includegraphics[width = 8.8 cm]{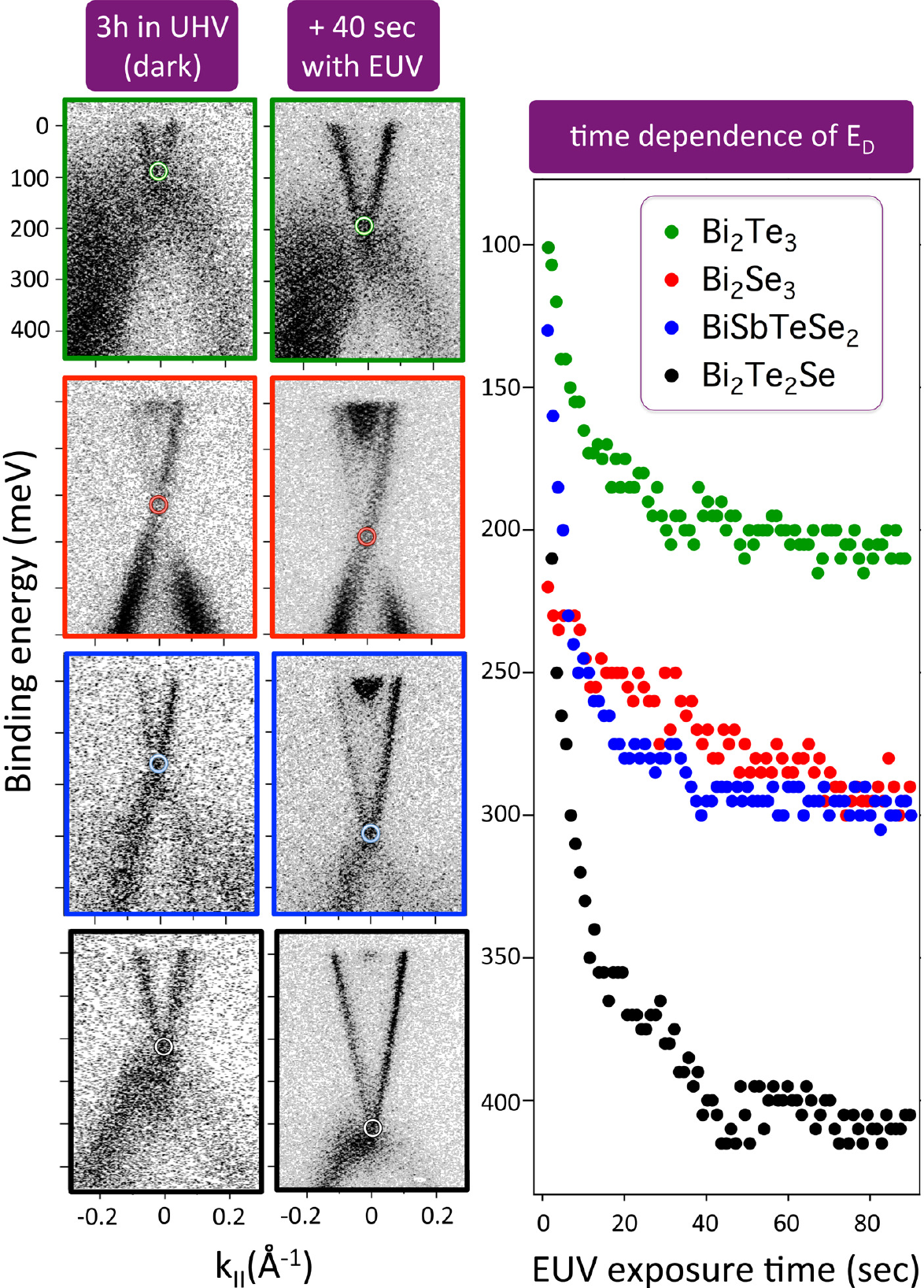}
  \caption{\textbf{The striking effect of illumination on the evolution of band bending during a typical ARPES experiment on TIs.} Left (center) columns of $I(E,k)$ images: surface band structure of four different TI compounds recorded using an EUV exposure of only one (40) second(s), after the cleavage surface had been exposed to residual UHV gases for 3 hours. Right column: very rapid evolution of $E_{\textmd{D}}$ from the first moment of EUV illumination, whereby the left-hand $I(E, k)$ images underlie the first (second for BiSbTeSe$_{2}$) data point of each curve in the right-hand panel. The pressure before and during the ARPES measurements was, $1.0 \times 10^{\textmd{-}10}$ and $5.0 \times 10^{\textmd{-}11}$ mbar, respectively. All data have been acquired at 16 K using a photon flux of $3.2 \times 10^{21}$ photons/(s m$^{2}$). The photon energy was 30 eV for Bi$_{2}$Se$_{3}$ and 27 eV for all the other compounds.}
\label{fig2}
\end{figure}

Fig. 2 shows the first experimental evidence that the surface band bending of 3D TIs is modified substantially on exposure to EUV illumination of a duration of a single second, compared to the typical timescale of ARPES data collection for an $I(E,k)$ image of tens of seconds or even several minutes. 
In order to highlight that the development of the band bending is indeed dominated by EUV exposure, and not by simple surface decoration with residual UHV gases, as has generally been believed \cite{Bianchi2010, Bianchi2011, King2011, Benia2011, Zhu2011, Bahramy2012, Chen2012, Bianchi2012}, we have constructed the following experimental protocol.
Firstly, we have intentionally exposed all cleavage surfaces to residual UHV gases for 3 hours at low temperature before the first measurement. Secondly, we have limited the duration of each measurement (and hence the EUV exposure) to a minimum of 1-2 seconds using a photon flux of $3.2 \times 10^{21}$ photons/(s m$^{2}$).
The optimization of the sample position with respect to the electron energy analyzer and the photon beam, and the adjustment of the emission angles -such that the detector image cuts through the center of the Brillouin zone- were carried out on a part of the cleave one or more millimeters away from the point where the data of Figs. 2 and 3 were recorded. This means that the $E_{\textmd{D}}$ values for the locations measured for Figs. 2 and 3 represent those for regions with carefully controlled EUV exposure \cite{Frantzeskakis2015-2}.    

Inspection of the green symbols in Fig. 1 - showing $E_{\textmd{D}}$ for EUV exposures of only 1-2 seconds - reveals that the Fermi energy is much closer to the Dirac point for each of the TI materials, compared to the red, open circles in Fig. 1 from other decorated surfaces which have not been intentionally `kept in the dark'  \cite{Bianchi2010, King2011, Bianchi2012, Arakane2012, Cao2014, Frantzeskakis2015}. 
This simply means that in the determination of $E_{\textmd{D}}$ using ARPES, \emph{every second counts}.

We have captured the rapid evolution of the surface band structure shown in the right-hand panel of Fig. 2, by keeping the sample in the dark, except for successive, 1-2 s short $I(E,k)$ ARPES exposures on the same spatial location.
The effect of EUV exposure is striking: the surface band structure changes within the very first seconds.
At the photon flux used, as little as 40 seconds EUV exposure yields the $I(E,k)$ images shown in the centre column of Fig. 2, and analysis of these data shows that $E_{\textmd{D}}$ is already approaching values typical for decorated surfaces. As a matter of fact, in terms of increasing the energy separation between $E_{\textmd{D}}$ and $E_{\textmd{F}}$, \emph{a few seconds of bright EUV light impacts the surface band structure of TI single crystals more than does an exposure to residual UHV gases in the dark in excess of a couple of hours.}

Aided by this new insight, we can assess the implications of these results for previous ARPES studies.
Even exploiting the high EUV photon flux of a 3$^{\textmd{rd}}$ generation synchrotron source, the typical acquisition time for an $I(E,k)$ image with a reasonable signal-to-noise ratio is of the order of 1-2 minutes. Moreover, the first measurement is usually preceded by the optimization of the spatial and angular positions of the sample; a process that usually takes place in the same sample location and the main measurements themselves.
This means that for the great majority of the published ARPES data, the sample surface has been exposed to EUV light for at least a few minutes before the acquisition of the first $I(E,k)$ band structure image at a specific location.
Given the speed of the EUV-triggered downward band bending on decorated (or aged) surfaces taking place within the first seconds of exposure, it is certainly feasible that this whole process can go unnoticed. There is therefore no surprise that changes in the electronic band structure have been solely ascribed to the adsorbed gases and no reference is usually made to the role of the radiation that is part of an ARPES experiment.

Returning to the discrepancy between ARPES and QO data on the size of the Fermi surfaces of the Dirac cone (and hence the energy difference between $E_{\textmd{D}}$ and $E_{\textmd{F}}$), it is clear that the huge effect of EUV light exposure lies at the core of the reported disagreement. Essentially, with the first few seconds of EUV exposure required to perform an ARPES measurement, the Dirac point shifts downward by more than 0.1 eV. Thus, due to its nature as an ionizing technique, ARPES sees a band-bent version of the band structure, different to that extracted from quantum oscillations in magneto-transport experiments. Logically, the longer the exposure to EUV radiation before the acquisition of the ARPES data measurement, the larger is the discrepancy between the two techniques, although the fact that a significant shift already takes place within the first half a minute should not be overlooked.
     
\begin{figure}
  \centering
  \includegraphics[width = 8.7 cm]{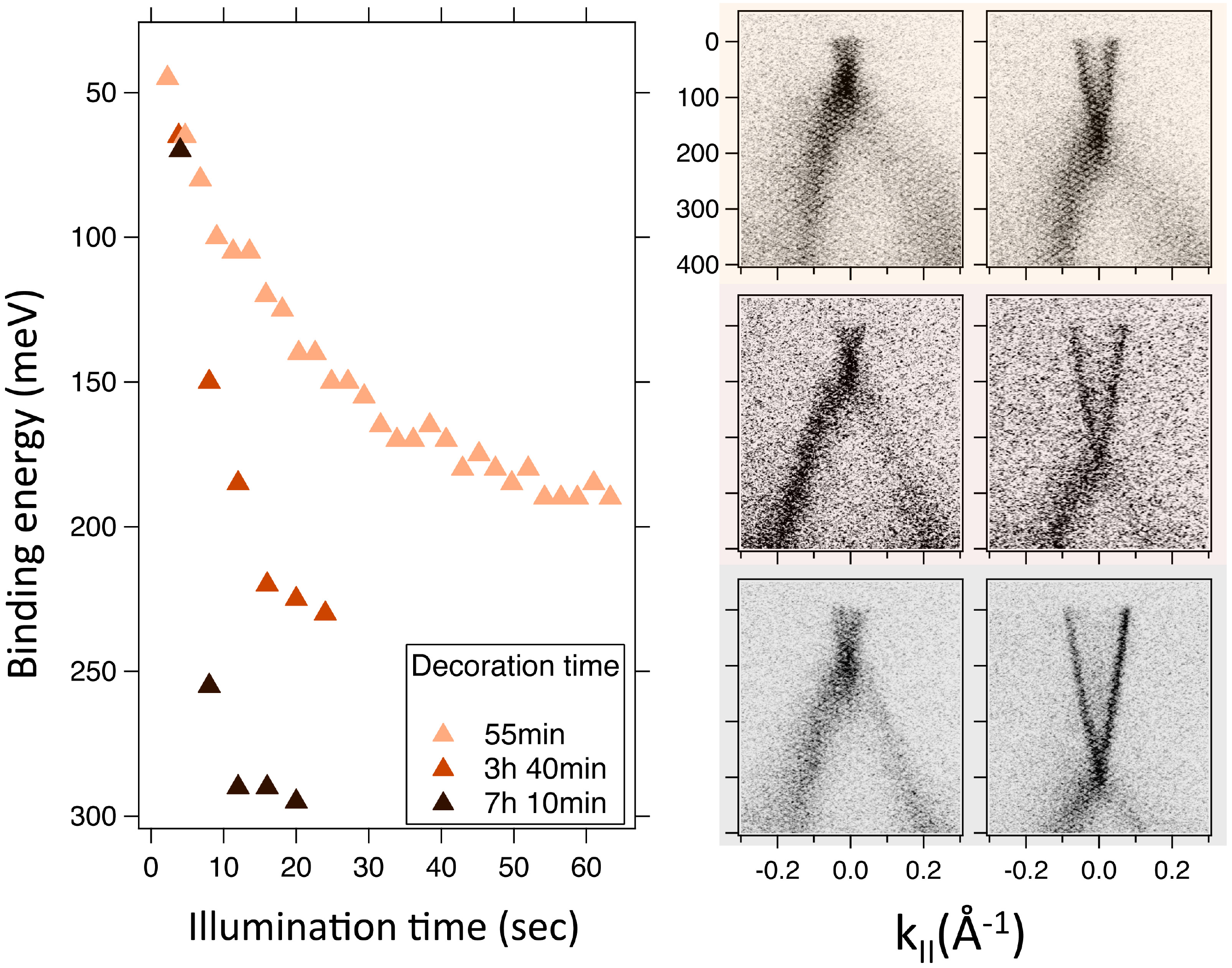}
  \caption{\textbf{The effect of surface decoration on the evolution of surface band bending during a typical ARPES experiment on 3D TIs.} Left column: binding energy of the Dirac point in Bi$_{1.46}$Sb$_{0.54}$Te$_{1.7}$Se$_{1.3}$ as a function of illumination time after the surface has been exposed for differing times to UHV residual gas molecules, as given in the legend. Center (right-hand) panels: the $I(E, k)$ images showing the surface band structure of Bi$_{1.46}$Sb$_{0.54}$Te$_{1.7}$Se$_{1.3}$ which underlie the data points in the left-hand graphic for EUV exposure times of one (20) s. The pressure before and during measurements was $1.0 \times 10^{\textmd{-}10}$ and $5.0 \times 10^{\textmd{-}11}$ mbar, respectively. All data have been acquired at 16 K using a photon flux of $3.2 \times 10^{21}$ photons/(s m$^{2}$), with a photon energy of 27 eV.}
\label{fig3}
\end{figure} 

From the data and discussion thus far, it is incontrovertible that EUV-triggering is a vital part of the band bending process.
However, further questions may naturally arise such as: (i) is the amount of surface decoration of importance in determining the evolution of the band bending? (ii) and can we extrapolate the ARPES results presented in Fig. 2 to $t\textmd{=}0$ in order to obtain $E_{\textmd{D}}$ under flat-band conditions? 

\begin{figure*}
  \centering
  \includegraphics[width =13.6 cm]{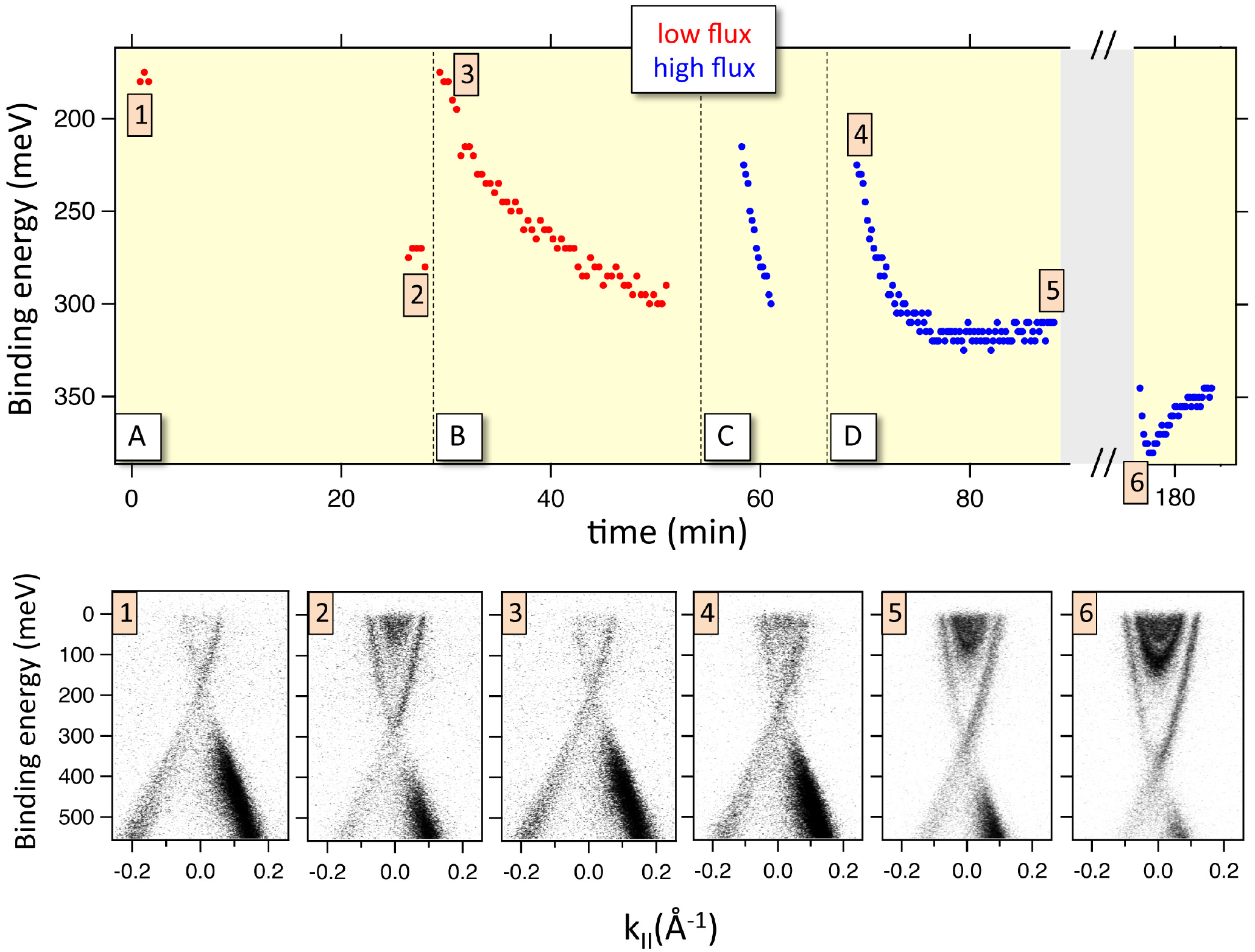}
  \caption{\textbf{The local character of EUV-induced surface band structure changes in 3D TIs, and the effect of photon flux.} The upper panels show the evolution of $E_{\textmd{D}}$ in Bi$_{2}$Se$_{3}$ on exposure of four different sample locations labelled $A$-$D$ to low (red) and high (blue) photon flux. Vertical dashed lines denote the change of sample location. The lower panels show the $I(E, k)$ images corresponding to the numbered points (1)-(6) superimposed on the curves in the uppermost panels. The pressure during measurements was $1.5 \times 10^{\textmd{-}10}$ mbar. The flux corresponding to the red data points was 3-4 times lower than the flux corresponding to the blue data points. All data have been acquired at 38 K using a photon energy of 30 eV.}
\label{fig4}
\end{figure*}

We can answer the first question positively and the second question negatively using the data presented in Fig. 3. The left-hand panel of the figure tracks the evolution of $E_{\textmd{D}}$ in three Bi$_{1.46}$Sb$_{0.54}$Te$_{1.7}$Se$_{1.3}$ (BSTS1.46) samples that have been measured using the same photon flux, but whose cleavage surfaces were held in residual UHV gases for differing time-periods before the first EUV exposure that is inevitably coupled to the ARPES acquisition process.
The differences between the time-dependence of $E_{\textmd{D}}$ in the three cases are marginal at the outset, and for short illumination times the $I(E,k)$ images look rather similar (see center column of Fig. 3). However, as the ARPES experiments and the ensuing EUV exposure progress, the sample with the longest a priori exposure to residual gases, and presumably the largest number of adatoms on its cleavage surface, exhibits a significantly more rapid evolution of its downward band bending.
Thus, already at time scales as short as 20 s total EUV exposure (right-hand column of Fig. 3), the three band structure images start to look quite different. This argues that the amount of surface decoration does still play a significant role, bearing great influence on both the saturation point and the evolution speed of the downward band bending. This said, the crucial role in the EUV light exposure in triggering the downward band bending should not be forgotten.
Consequently, it is evident that we cannot simply extrapolate the curves of either Fig. 2 (3 hours UHV residual gas) or Fig. 3 ($\sim$1, $\sim$4 and 7h UHV residual gas) to $t\textmd{=}0$, in an attempt to extract the energy difference between $E_{\textmd{F}}$ and $E_{\textmd{D}}$ under flat-band conditions, as such a procedure would ignore the differences in the UHV `decoration time' from when the cleavage surface was in the dark prior to the first illumination.
Nevertheless, what Figs. 2 and 3 show is that as long as the ARPES acquisition time is kept very rapid - meaning well below 5 s - the very first ARPES measurement can provide us with an upper limit of the energy separation of $E_{\textmd{F}}$ and $E_{\textmd{D}}$ for flat-band conditions.
We note that the Fermi level (flat-band conditions) should be pinned close to the conduction band minimum (CBM) and the valence band maximum (VBM) for $n$-type and $p$-type samples, respectively \cite{Brahlek2015}.
Past studies have identified $n$-type carriers in Bi$_{2}$Se$_{3}$ and Bi$_{2}$Te$_{2}$Se \cite{Cava2013, Ren2010, Analytis2010, Ren2011, Xiong2012, Xiong2012-2, Cao2014, Frantzeskakis2015}, and $p$-type carriers in Bi$_{2}$Te$_{3}$, BiSbTeSe$_{2}$ and other compounds of the Bi$_{2\textmd{-x}}$Sb$_{\textmd{x}}$Te$_{3\textmd{-y}}$Se$_{\textmd{y}}$ family \cite{Cava2013, Qu2010, Ren2011, Frantzeskakis2015}.
The initial $I(E,k)$ images - recorded within 5 seconds of the onset of illumination - for the five TI compounds whose data are shown in Figs. 2 and 3 are in good agreement with these previous assignments, with the VBM of the $p$-type compounds lying between 50-130 meV below the Fermi level, while its binding energy exceeds 200 meV for all of the $n$-type compounds. 

Fig. 3 illustrates that the speed of development of the downward band bending increases as the degree of surface decoration prior to illumination increases.
Fig. 4 shows that this degree of decoration is not the only parameter, and that variations of the photon flux from one experiment to the next can also accelerate or decelerate the evolution of the downward band bending.
The upper panels of Fig. 4 tracks the Dirac point energy over time in a sample of Bi$_{2}$Se$_{3}$ which - in differing locations - has been exposed to both low (red data points, panels $A$ and $B$) and high photon fluxes (blue data points, panels $C$ and $D$). The reduced photon flux was obtained by detuning the undulator gap with respect to the ideal value for the photon energy being selected by the monochromator, and from the ARPES signal intensity, we conclude that the attenuation was by a factor of 3-4.
At the beamline/endstation combination used for the experiments underlying Fig. 4, acquisition as fast as that used for Figs. 2-3 would have yielded insufficient signal/noise, and thus the focus was placed on the behaviour over the longer time scales of minutes, rather than seconds.   
Experiments were performed at four different sample locations (labelled $A$, $B$, $C$ and $D$ in Fig. 4), each sufficiently well separated from the others so that each location enables an independent experiment.
One can readily see that the downward time evolution of $E_{\textmd{D}}$ is more rapid when the photon flux is higher (blue data points in panels $C$ and $D$).
Turning to the lower panels of Fig. 4, which show $I(E, k)$ images recorded along the timer/location series as shown in the upper panels of the figure, a comparison of the $I(E, k)$ images (1), (2) and (3) shows that the surface band structure changes are essentially local in nature, as $E_{\textmd{D}}$ goes back to close to its initial value when a previously unirradiated sample location is measured. 
For the start of the high-flux acquisition series shown in panel (4) of Fig. 4, we see that $E_{\textmd{D}}$ is at slightly higher binding energy than the other `fresh' scans [(1) and (3)]. We ascribe this difference to a combination of the very fast initial evolution of the surface band structure under high flux illumination that cannot be captured in this case, and a slow but finite spreading out of local surface band structure changes away from the illuminated location.
The local character of photo-induced surface band structure changes has been discussed in detail in Ref. \onlinecite{Frantzeskakis2015-2}, and this was used to write micrometric structures in which topologically trivial electronic states could be pushed above the Fermi level in pre-defined areas.

After the downward bend bending affecting the surface band structure in location $D$ had saturated [panel (5) of Fig. 4], the EUV exposure was halted for 90 min., while leaving the sample at the measurement position (see the grey area in the right-hand top panel of the figure).
On subsequent continuation of the EUV exposure, the downward band bending is seen to grow rapidly once again, thereby shifting $E_{\textmd{D}}$ to an even higher value of binding energy as shown in panel (6). Finally, as the exposure of the sample to high flux radiation continues, $E_{\textmd{D}}$ starts shifting back to lower values, in a manner first presented and described in detail elsewhere \cite{Frantzeskakis2015-2}.   

Taking the data presented in Figs. 1-4 together, we have shown that there is a persistent discrepancy between the Dirac point energies extracted from published ARPES data and those from Fermi surface analyses based on quantum oscillation experiments.
We showed that the EUV exposure inherent to the ARPES experiment itself triggers the downward band bending, and that this lies at the root of the discrepancy with the transport data.
We also showed that both the initial surface decoration with adatoms from the UHV residual gases, and the photon flux used to carry out the photoemission experiment are parameters that influence the speed with which the downward band bending evolves in common Bi-based 3D TI compounds.
ARPES is a central experimental technique in the TI field, but the data indicate significant changes in the energy separation between the Fermi and Dirac energies on the timescale of mere seconds after the very first exposure to EUV light.
Therefore, in the following section we show the results of experiments designed to reveal the complete evolution of the surface band structure for a representative 3D topological insulator during a standard ARPES experiment at a large-scale facility, covering both the downward band-bending and subsequent upward shift of the bands. We discuss a simplified mechanism that can explain these changes. \\

\section{Discussion}

It is clear by now that determining the energy of the Dirac point in topological insulators by means of a photon-based experimental technique like ARPES is far from trivial.
If the samples enable sufficient suppression of the bulk conductivity, transport studies possess a clear advantage: even though the surface is certainly exposed to external gas molecules, these do not affect its surface electronic band structure because the trigger for downward band bending active in ARPES - the strong EUV illumination - is missing.
On the other hand, ARPES is very sensitive to the surface states, and provides a direct access to the $E$ vs. $k$ dispersion relation, something which is of great value.
Thus, despite the complications brought with the EUV illumination `built into' ARPES, it is important to complete the discussion on how to understand the results expected in a standard, low temperature ARPES experiment on 3D TI surface states.
In the following paragraphs, we shall focus on the mechanism behind the time-dependent evolution of the band structure, and suggest two alternative routes that allow flat-band conditions to be approached and hence the `real' energy value of the Dirac point to be determined using ARPES.

\begin{figure*}
  \centering
  \includegraphics[width = 16.0 cm]{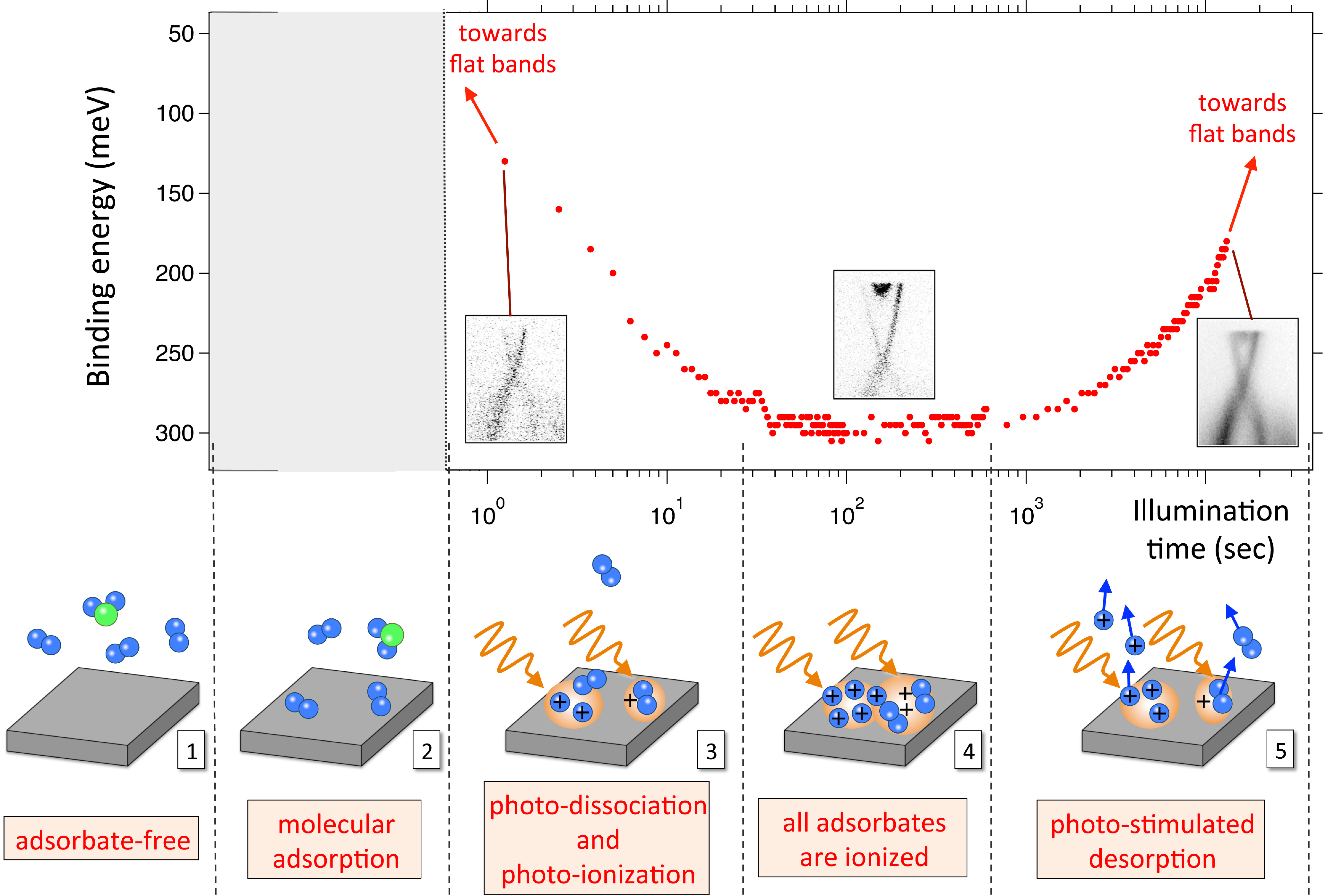}
  \caption{\textbf{The complete evolution of the band structure of TIs under intense illumination and a schematic of the underlying mechanism.} (Top panel) Evolution of $E_{\textmd{D}}$ at different timescales obtained on a single surface location of BiSbTeSe$_{2}$ during a standard ARPES experiment. From left to right: intense illumination promotes downward band bending until a saturation point and, from then on further exposure tends to flatten the bands again. Red arrows indicate that true flat-band conditions in an ARPES experiment on these materials can only be achieved at very short or very long timescales. The insets in the upper panel show the corresponding $I(E, k)$ diagrams at different illumination times. Prior to illuminating the surface location that was finally studied with ARPES, the sample had been exposed to residual gas molecules for 3 hours, represented by the grey shaded area in the upper panel. The base pressure before and during measurements was, respectively, $1.0 \times 10^{\textmd{-}10}$ and $5.0 \times 10^{\textmd{-}11}$ mbar. All data have been acquired at 16 K using a 27 eV photon flux of $3.2 \times 10^{21}$ photons/(s m$^{2}$). (Bottom panels) Schematics of the five steps underlying our observations. Vertical dashed lines denote the time period to which each step corresponds. Step 1: immediately after cleavage, the sample surface is free of adsorbates. Step 2: residual gas molecules adsorb on the sample surface during the period while the latter is held in the `dark'. Step 3: under the influence of the intense EUV photon beam required to do ARPES, adsorbed molecules start to dissociate into single adatoms in the illuminated sample location \cite{Jiang2012}. Step 4: all adsorbed molecules are dissociated and/or photo-ionized, and band bending reaches a maximum. Step 5: Photo-stimulated desorption of adatoms and surface photovoltage (not illustrated) come into play and the bands shift back up again.}
\label{fig4}
\end{figure*}

Our results have established that the evolution of the surface band structure of common Bi-based TIs during an ARPES experiment -and more generally under any experiment involving intense (E)UV illumination- is not random.
It is determined by the initial surface decoration (as shown in Fig. 3) and the magnitude of the photon flux (as shown in Fig. 4). However, the whole band bending process is only triggered by the onset of the EUV illumination itself (as shown in Fig. 2).
The interplay between these parameters results in a characteristic time-dependent behavior of $E_{\textmd{D}}$, an example of which is shown in the upper panel of Fig. 5. Parts of this time-dependent process have certainly been discussed in the framework of the earlier figures in this paper and in a number of prior publications \cite{Bianchi2010, Benia2011, Zhu2011, King2011, Bahramy2012, Jiang2012, Frantzeskakis2015}, yet the data of Fig. 5 present the first comprehensive description of \emph{the complete temporal evolution spanning from the very first seconds to several hours of continuous EUV illumination} on the same sample location.

The data points in the upper panel of Fig. 5 are actually from a cleaved crystal of the bulk-insulating 3D TI BiSbTeSe$_{2}$ which has been exposed to residual UHV gases for 3 hours before starting illumination. The logarithmic timescale chosen for the x-axis of Fig. 5 stresses the importance of the first seconds of light exposure. 
The $I(E,k)$ image insets illustrate the surface band structure.
At first (left-most image), the Fermi level lies 130 meV above the Dirac point, and no bulk conduction band states are seen below the Fermi level.
As time progresses, $E_{\textmd{D}}$ continuously increases, saturating at 300 meV below the Fermi energy (center inset image).
With the photon flux used for these data of $3.2 \times 10^{21}$ photons/(s m$^{2}$), this saturation is achieved in less than 5 min., and within this timescale, the bulk conduction band has been shifted into the occupied part of the energy spectrum.
On further EUV exposure, the trend is inverted, whereby $E_{\textmd{D}}$ then starts to shift back towards $E_{\textmd{F}}$, within 3-4 hours reaching an energy value that is comparable to that after the first second of illumination (right-most inset).
The observed spectral broadening and the trace of a conduction band tail are the result of the inhomogeneous intensity profile of the beam spot, and this phenomenon has been the focus of previous investigations \cite{Frantzeskakis2015, Frantzeskakis2015-2}.
Here we stress the opposite trends of the two photo-induced phenomena at small and large timescales, as well as the remarkable symmetry of the logarithmic time-dependent curve around the saturation point. 

Thus, we are now in a position to give a complete answer to the question of how we can approach flat-band conditions in an ARPES experiment on such topological insulators.
For a photon flux of $3.2 \times 10^{21}$ photons/(s m$^{2}$), which is of the order of the standard photon flux at a 3$^{\textmd{rd}}$ generation synchrotron facility, downward band bending is reduced to a minimum at illumination times either smaller than one second, or exceeding 4-5 hours.
We have previously shown that approaching flat-band conditions at large timescales is only possible at high photon fluxes and energies \cite{Frantzeskakis2015}, and that the data are always subject to spectral broadening \cite{Frantzeskakis2015, Frantzeskakis2015-2}.
Therefore, very short illumination times represent the preferred option to approach flat-band conditions.
We note that by reducing the flux, the initial, downward band bending can be decelerated, so as to fit such a `fast' measurement into the practical time window of a given ARPES set-up, thus yielding band structures that are even closer to flat bands (see Fig. 4).
ARPES experiments at very small or very large timescales are hence a prerequisite for meaningful comparison with the results from transport.
Indeed, the ARPES data points from the very first illumination on each sample in this paper (shown as green markers in Fig. 1) are in reasonable agreement with the SdH data.

We can now turn to the origin of the observed time dependence of the electronic band structure.
Cartoons in the bottom row of Fig. 5 illustrate a sequence of microscopic processes underlying each experimental step.
Right after cleavage, the surface is pristine and there is no band bending (panel 1).
While it ages in the dark, the cleavage surface is decorated with residual gas molecules (panel 2).
Specifically, molecular hydrogen - as the dominant residual gas in clean UHV vacuum systems - is a possible adsorbate.
It has been proposed to photo-dissociate, resulting in stronger bonding with the sample surface, and donation of electron density to the TI \cite{Jiang2012}.
Also conceivable is the photo-ionization of adsorbed molecules. Both processes (photo-dissociation and photo-ionization) result in a positive ion layer developing at the surface-vacuum interface, resulting in downward band bending \cite{Bianchi2010, Benia2011, Zhu2011, King2011, Bahramy2012, Jiang2012} (panel 3).
When all adsorbed molecules in the illuminated area have either dissociated (in the case of H$_{2}$) or in any case have become ionized, the binding energy of the Dirac point and the downward band bending reach a maximum (panel 4). Nevertheless, the role of illumination is not limited to dissociation/ionization of molecular species: irradiation may also promote the desorption of adatoms and molecules. Photo-stimulated desorption plays a role in the observed flattening of the bands with $E_{\textmd{D}}$ moving back to lower binding energies seen in panel 5.
In addition, an additional photo-induced electronic phenomenon, known as the surface photovoltage (SPV) effect, creates electron-hole pairs in the space-charge region that act in synergy with photo-stimulated desorption to further decrease the downward band bending \cite{Frantzeskakis2015}.

The mechanism based on the combination of molecular adsorption, photo-dissociation, ionization, photo-stimulated desorption and SPV can explain our observations and the results of previous studies \cite{Bianchi2010, King2011, Benia2011, Zhu2011, Bahramy2012, Bianchi2012, Jiang2012}.
In a typical ARPES experiment, the photon beam does not stay on the same sample location long enough to observe the relatively slow upward shift of $E_{\textmd{D}}$ seen in the right hand side of Fig. 5.
Most of past literature has been therefore focused on the development of the initial downward band bending that is governed by molecular adsorption, photo-dissociation and/or ionization.
The latter two processes are relatively fast, and thus can remain unnoticed as the overall dynamics would then be governed by the slow molecular adsorption step. By intentionally increasing the time of exposure to residual gas atoms before the beginning of irradiation, we separate photo-dissociation/ionization from molecular adsorption.
While adsorption increases during aging in the dark, there are no photons to generate the charged surface layer.
Thus, immediately upon the onset of illumination, the effects of photo-dissociation/ionization become clear.
Our strategy to experimentally separate the two processes and to decrease the time window of the measurements themselves so as to better match the more rapid dynamics of photo-dissociation/ionization has revealed that illumination is the essential trigger of the time-dependent band structure changes in Bi-based TIs.
This result adds important nuance to the common belief that adsorption is the sole process at the origin of band bending.
 
Can this new insight be tested?
Returning to Fig. 4, the time-dependent curve shown at location $D$ in the upper panel can now be re-examined. On saturation of the downward band bending [point (5) / panel (5)], a steady state is reached between dissociation/ionization of adsorbed molecules on one hand, and photo-stimulated desorption/SPV on the other hand.
While the sample is then kept further in the dark, the latter effects are `turned off' and molecular adsorption continues.
Upon resuming irradiation, photo-dissociation and ionization of the newly adsorbed molecules comes back into play, governing  the first seconds of illumination and leading to a renewed increase in the downward band bending.
At later times, the upturn in the Dirac point energy suggests that photo-stimulated desorption is important in re-asserting the flattening of the bands.
Summarizing this discussion, the band structure dynamics in the ARPES of Bi-based TIs are governed by photo-dissociation/ionization at short timescales, by molecular adsorption at timescales between a very few minutes and the saturation of downward band bending, and by photo-stimulated desorption and SPV at later times.
It is the ensemble of these processes - all of which are inherent to an ARPES experiment conducted at low temperature - that is the reason for the discrepancies seen between ARPES and magneto-transport experiments as regards the electronic band structure of Bi-based 3D TIs.\\

\section{Conclusions}

First of all, at a fundamental level, we have proposed and discussed a sequence of processes that lie behind the complex evolution of the electronic band structure of Bi-based 3D TIs under intense (E)UV illumination.
The interplay of molecular adsorption, photo-dissociation, ionization, photo-stimulated adsorption and surface photovoltage are able to account for the observed changes during a typical experiment carried out by means of photon-based techniques such as ARPES.
The dynamics of these underlying processes differ significantly, and hence govern the band structure evolution at different timescales.

Secondly, at a practical level, we have underlined all parameters that need to be taken into account so as to predict the details of the band structure during an ARPES experiment on Bi-based TIs. Most notably, we have determined two alternatives in order to approach flat-band conditions with ARPES: measurements after very short or very long illumination times.

Furthermore, we have cleared up the apparent discrepancies as regards the energy position of the Dirac point from ARPES and SdH oscillations. Therefore, with a view towards possible technological applications based on TIs, our results will facilitate the choice of the most suitable compound.

Last but not least, the established local character of the initial photo-induced changes permits the selection of sample areas where downward band bending is induced. In other words, one may define channels of a combined trivial and topological conduction within a purely topological matrix (by photo-triggering the band bending), as a twin to the erasure of the trivial states from the combined system via photo-stimulated desorption or via surface photovoltage. Such effects might find applications in emerging devices based on topological transport.\\

\section{Methods}

\subsection*{Sample growth} Crystals were grown in Amsterdam using the Bridgman technique. High purity elements were melted in evacuated, sealed quartz tubes at 850$^{\textmd{o}}$C and allowed to mix for 24 hours before cooling.
The cooling rate was 3$^{\textmd{o}}$C per hour. Samples were cleaved and aged in UHV ($P<5.0\times10^{\textmd{-}10}$ mbar) and at temperatures ranging from 16 to 38 K.\\

\subsection*{Angle resolved photoelectron spectroscopy measurements} ARPES experiments were performed at two different experimental setups.\\
(i) At the SIS-HRPES endstation of the Swiss Light Source with a Scienta R4000 hemispherical electron analyzer. The minimum temperature was 16 K and the pressure during measurements was $5.0\times10^{\textmd{-}11}$ mbar.\\
(ii) At the UE112-PGM-2a-1$^{2}$ endstation of BESSY II with a Scienta R8000 hemispherical electron analyzer. The minimum temperature was 38 K and the pressure during measurements was $1.5\times10^{\textmd{-}10}$ mbar.\\
ARPES data were measured using 27 eV photons (or 30 eV for Bi$_{2}$Se$_{3}$). The polarization was linear horizontal except for data presented in Fig. 4 where we used circular polarization. To determine the effect of photon flux, the latter was deliberately decreased by detuning the undulator energy, while the nominal monochromator energy remained fixed. Relative values of the photon flux were deduced by comparing the ARPES intensity under the same experimental conditions. The duration of each ARPES measurement during the initial evolution of surface band bending was between 1 and 6 s. An accurate determination of the position of the Fermi level was determined from measurements on an in-situ evaporated Au thin film held in electrical contact with the sample. \\

\section{Appendix: A model of the microscopic mechanisms}
\setcounter{figure}{0}
\renewcommand{\thefigure}{A\arabic{figure}}

In the appendix we will use model equations to arrive at a quantitative understanding of the phenomena described in the main text. Using simple formulas we will simulate the microscopic processes of molecular adsorption, photo-induced dissociation/ionization and photo-stimulated desorption. Finally, we will combine the effects of those processes on the energy shift of the surface electronic structure to arrive at a quantitative description of the evolution of surface band bending. 
Fig. 4 will serve as experimental input to describe the time-dependent band bending observed in the surface electronic band structure of Bi$_2$Se$_3$.

Following Refs. \onlinecite{Frantzeskakis2015, Analytis2010,Brahlek2015,Cava2013}, and the discussion in the frame of Fig. 2 of the main text, we assume that, at flat band conditions, the Fermi level of Bi$_2$Se$_3$ is pinned at the bottom of the conduction band. On exposure to adsorbates and illumination, $n$-type band bending develops and the binding energy of the Dirac point ($E_{\textmd{D}}$) increases. The uppermost data point in Fig. 4 serves as an approximation for $E_{\textmd{D}}$ at flat band conditions. The surface carrier density $N_{\textmd{s}}$ that corresponds to a Fermi level position at 175 meV above $E_{\textmd{D}}$ can be expressed as:
\begin{equation}
N_\textmd{s}=\frac{k_\textmd{F}^{2}}{2\pi}  \quad \textmd{and} \quad E_\textmd{F}=\hbar\cdot\textsl{v}_\textmd{F} \cdot k_\textmd{F} 
\end{equation}\\       
where $E_\textmd{F}$ denotes the energy difference between the Fermi level and the Dirac point (i.e 175 meV), $k_\textmd{F}$ is the Fermi wavevector, $\textsl{v}_\textmd{F}$ the Fermi velocity and $\hbar$ Planck's constant. The Fermi velocity of Bi$_2$Se$_3$ is $5.3\times10^{5}$ m/s (Fig. 4) and the equations yield an intrinsic surface carrier density of $4.0\times10^{12}$ cm$^{\textmd{-}2}$ at flat band conditions. This value is in good agreement to what has been previously reported by Brahlek et al. in Ref. \onlinecite{Brahlek2015}. We can calculate the extra charges that lead to band bending by the difference of the surface carrier density at maximum band bending and under flat band conditions. We assume that panel (6) of Fig. 4 shows the ultimate band bending conditions of Bi$_2$Se$_3$ and we repeat the aforementioned procedure to find $N_\textmd{s}=1.89\times10^{13}$ cm$^{\textmd{-}2}$. The maximum density of the adsorbate-induced surface carriers is therefore equal to $1.49\times10^{13}$ cm$^{\textmd{-}2}$. This density corresponds to $1$ charge per $6.71$ nm$^2$ and to a coverage of $0.022$ adsorbate molecules per substrate atom, under the assumption that we have a single charge per adsorbate and after taking into account the in-plane lattice parameter of Bi$_2$Se$_3$.\\

\subsection*{Dynamics of molecular adsorption}
At low coverages the sticking coefficient ($s$) can be considered constant. The time dependence of the molecular coverage is then given by a simple linear equation:
\begin{equation}
\Theta_{\textmd{mol}}(t)=s\cdot Z\cdot t 
\end{equation} \\                     
where Z, the arrival rate of molecules (collision flux), is given by:
\begin{equation}
Z (\textmd{cm}^{\textmd{-}2}\textmd{s}^{\textmd{-}1})=2.63\times10^{22}\cdot \frac{P}{\sqrt{M\cdot T}}                    
\end{equation}\\
with $P$ denoting the pressure (mbar), T the temperature (K) and $M$ the molecular mass (g/mol). Using the values of $P$ ($1.5\times10^{\textmd{-}10}$ mbar) and $T$ (38 K) from the caption of Fig. 4, an average molecular mass of 29 g/mol for air molecules and the surface density of substrate atoms, we can rewrite equation (2) in terms of molecular coverage per substrate atom:
\begin{equation}
\Theta_{\textmd{mol}}(t)=1.77\times10^{\textmd{-}4}\cdot s\cdot t                         
\end{equation}

For $s=1$, we need 124 seconds to obtain a molecular coverage that corresponds to the maximum charge coverage of $0.022$. Although there is no previous reference on the sticking probability of air molecules on Bi$_2$Se$_3$, the sticking coefficient of the most important contaminant in UHV, molecular hydrogen, has been reported to be much smaller than unity on several different surfaces \cite{Winkler1982, Berger1992, Winkler1998, Kraus2016}. In order to get an experimental estimate of the sticking probability from the data presented in Fig. 4, we compare the binding energy of the Dirac point in panels (5) and (6), i.e. before and after 90 minutes in the dark. The molecules that arrived in the surface during this time period were later activated by the light beam and drove the Fermi level 65 meV higher. The corresponding increase in surface density of adsorbate-induced carriers is equal to $5.9\times10^{12}\textmd{cm}^{\textmd{-}2}$. A constant adsorption rate would mean that the coverage of $0.022$ adsorbate molecules per substrate atom would be reached in 225 minutes. A comparison of this value with the value of 124 seconds obtained assuming a sticking coefficient of unity, yields a sticking coefficient $s$ of the order of $0.01$. This result is within the range of the sticking coefficient of molecular hydrogen in the aforementioned past studies \cite{Winkler1982, Berger1992, Winkler1998, Kraus2016}.

The corresponding time-dependent density of adsorbed molecules can be expressed as:
\begin{equation}
N_{\textmd{mol}}(t)=1.19\times10^{15}\cdot s\cdot t
\end{equation}\\
where the value corresponding maximum charge density of  $1.49\times10^{13} \textmd{cm}^{\textmd{-}2} = 1.49\times10^{17} \textmd{m}^{\textmd{-}2}$ is reached in 124 seconds for $s=1$.  

We underline that the maximum charge coverage of $0.022$ does not mean that molecular coverage saturates at this value. Charges are continuously being created by photo-induced dissociation and ionization while they are at the same time destroyed by photo-stimulated desorption. In order to achieve a maximum charge coverage of $0.022$, one needs to take into account the charges lost due to desorption and this is only possible if the supply of molecules does not saturate at $0.022$ but at a higher value. We will note this last value as $\Theta_{\textmd{mol} \, \textmd{max}}$ and we will use it as a partially-free parameter in our model with a lower limit of $0.022$.\\

\subsection*{Dynamics of photo-induced dissociation/ionization}
We assume that adsorbed molecules do not yield any charges before their activation with the incoming photons either via photo-induced dissociation ($\textmd{=}$ photolysis) or photo-ionization. We further assume that the dynamics of these processes are identical and they are determined by the absorption of photons. The first step is to calculate the absorption rate of photons ($R$). 
\begin{equation}
R=F\cdot \sigma_{a}                    
\end{equation}\\
where $F$ is the photon flux in no. of photons / (s$\cdot$ m$^{2}$) and $\sigma_{a}$ is the cross section for photon absorption. In the following, we will use the photo-absorption cross-section of molecular hydrogen at 30 eV ($\sigma_{a}=2.6\times10^{-22}$ m$^2$) and we will consider the photon flux as a partially-free parameter in our model. The upper limit of the photon flux is $3.2\times10^{21}$ photons/(s$\cdot$ m$^{2}$) (flux used in Figs. 1-3 and 5) but we note that the flux used in Fig. 4 was unknown but considerably lower.

Using equation (5) we can now express the rate of absorption events ($r$) as:
\begin{equation}
\begin{split}
r= R\cdot \Theta_{\textmd{mol}}(t)&=F\cdot \sigma_{a}\cdot \Theta_{\textmd{mol}}(t) \\
                                                      &=F\cdot2.6\times10^{\textmd{-}22}\times1.77\times10^{\textmd{-}4}\cdot s\cdot t \\ 
                                                      &=F\cdot s\cdot4.6\times10^{\textmd{-}26}\cdot t
\end{split}
\end{equation}\\
where $r$ is expressed in no. of absorption events/sec, $s$ is the sticking coefficient of the order of 0.01 and $F$ is the photon flux with an upper limit of $3.2\times10^{21}$ photons/(s$\cdot$ m$^{2}$). 

We simulate photolysis and photo-ionization as an exponential process with $r$ giving the rate constant. The density of the created charges and the corresponding charge coverage can be written as:
\begin{equation}
N_{\textmd{charge} \, \textmd{creation}}(t)=N_{\textmd{mol}}(t)\cdot\left(1\textmd{-}e^{\textmd{-}r\cdot t}\right)                                                   
\end{equation}
\begin{equation}
\Theta_{\textmd{charge} \, \textmd{creation}}(t)=\Theta_{\textmd{mol}}(t)\cdot\left(1\textmd{-}e^{\textmd{-}r\cdot t}\right)
\end{equation}\\
$N_{\textmd{charge} \, \textmd{creation}}$ and $\Theta_{\textmd{charge} \, \textmd{creation}}$ are limited by the molecular adsorption and thus they can never exceed the corresponding values of $N_{\textmd{mol}}$ and $\Theta_{\textmd{mol}}$. They will however reach values much higher than the calculated maximum values of $\Theta_{\textmd{charge} \, \textmd{tot}}=0.022$ and $N_{\textmd{charge} \, \textmd{tot}}=1.49\times10^{17}$ m$^{\textmd{-}2}$ as charge losses due to photo-stimulated desorption and surface photovoltage (SPV) are not taken yet into account.\\

\subsection*{Dynamics of photo-stimulated desorption}
In the present subsection, we model the desorption of surface gas molecules due to the absorption of incoming photons. The removal of these molecules before being activated by photons through photolysis and photo-ionization means that they cannot yield surface carriers. The process of surface photovoltage that may also result in charge loss is not considered in the following dynamics. The rate of absorption events has already been presented in equation (7). If we assume a first-order desorption process the time-dependent removal of molecules, in terms of molecular density and coverage, is given by simple exponentials:
\begin{equation}
N_{\textmd{mol} \, \textmd{loss}}(t)=N_{\textmd{mol}}(t)\cdot e^{\textmd{-}r\cdot t}                                                   
\end{equation}
\begin{equation}
\Theta_{\textmd{mol} \, \textmd{loss}}(t)=\Theta_{\textmd{mol}}(t)\cdot e^{\textmd{-}r\cdot t}
\end{equation}\\
where those expressions start from a maximum value of molecular density and coverage for $t=0$ and tend exponentially towards zero carriers as time increases. These equations take into account the process of molecular adsorption but they do not consider charge creation via photolysis and photo-ionization.\\

\subsection*{Total dynamics}
The challenge now is to combine the dynamics expressed in equations (8)-(11) in a single formula that takes into account both the creation and loss of charges. For the sake of simplicity we further assume that photo-stimulated desorption applies only to molecules that have been already activated via photolysis or photo-ionization. In other words, the first photo-absorption event by an adsorbed molecule always results in its activation via photolysis or ionization: a molecule can be desorbed only after it has been dissociated or ionized. Under this assumption we can express the total dynamics as follows:
\begin{equation}
N_{\textmd{charge} \, \textmd{tot}}(t)=N_{\textmd{charge} \, \textmd{creation}}(t)\cdot e^{\textmd{-}r\cdot t}                                                   
\end{equation}
\begin{equation}
\Theta_{\textmd{charge} \, \textmd{tot}}(t)=\Theta_{\textmd{charge} \, \textmd{creation}}(t)\cdot e^{\textmd{-}r\cdot t}
\end{equation}\\
One can readily see that charge loss is now ultimately connected to charge creation which in turns depends on molecular adsorption via equation (8) and (9).

The final time-dependence of the Fermi level energy can be calculated using equations (1) where the carrier density will be given from (12):
\begin{equation}
E_{\textmd{F}}(t)=\sqrt{2\pi(N_{\textmd{charge} \, \textmd{tot}}(t)+ 4.0\times10^{16})}\cdot \hbar\cdot \textsl{v}_\textmd{F}                                                             
\end{equation}

We note that the intrinsic charges, already introduced in the first subsection of the appendix, have to be added to the adsorbate-induced charges for the calculation of the Fermi level energy. We recall that the intrinsic carrier density is equal to $4.0\times10^{16}$ m$^{\textmd{-}2}$.

Before plotting the model results, we will make two final modifications in the time-dependent equations that describe molecular adsorption [equations (4) and (5)]. First of all, molecular adsorption starts already during the time `in the dark'. Therefore, if $t=0$ marks the beginning of illumination equations (4) and (5) should become:
\begin{equation}
\Theta_{\textmd{mol}}(t)=1.77\times10^{\textmd{-}4}\cdot s\cdot(t+t_{\textmd{d}})                                                   
\end{equation}
\begin{equation}
N_{\textmd{mol}}(t)=1.19\times10^{15}\cdot s\cdot(t+t_{\textmd{d}})
\end{equation}\\
where $t_{\textmd{d}}$ denotes the time interval (in seconds) before the first photons arrive on the specific sample location. In the data of Fig. 4, $t_{\textmd{d}}$ is of the order of 1.5-2 hours for panel (4) and of 3-3.5 hours for panel (6). 

The last modification is related to a saturation coverage of the adsorbed molecules. Equations (15) and (16) describe a molecular coverage and density that increase without a limit. This is physically unreasonable. We will therefore make the assumption that both $\Theta_{\textmd{mol}}(t)$ and $N_{\textmd{mol}}(t)$ reach a plateau at $\Theta_{\textmd{mol} \, \textmd{max}}$ and $N_{\textmd{mol} \, \textmd{max}}$. These values will be considered as a partially free parameter in our model with a lower limit at $0.022$ for $\Theta_{\textmd{mol} \, \textmd{max}}$ and at $1.49\times10^{13}$ cm$^{\textmd{-}2}$ for $N_{\textmd{mol} \, \textmd{max}}$. Provided that $t_{\textmd{d}}$ is of the order of several minutes or larger (typical for an ARPES experiment), the exact values of the saturation coverage and saturation density modify only slightly the time-dependent quantities expressed by equations (12)-(14).\\

\subsection*{Model results}
\begin{figure}
  \centering
  \includegraphics[width = 8.2cm]{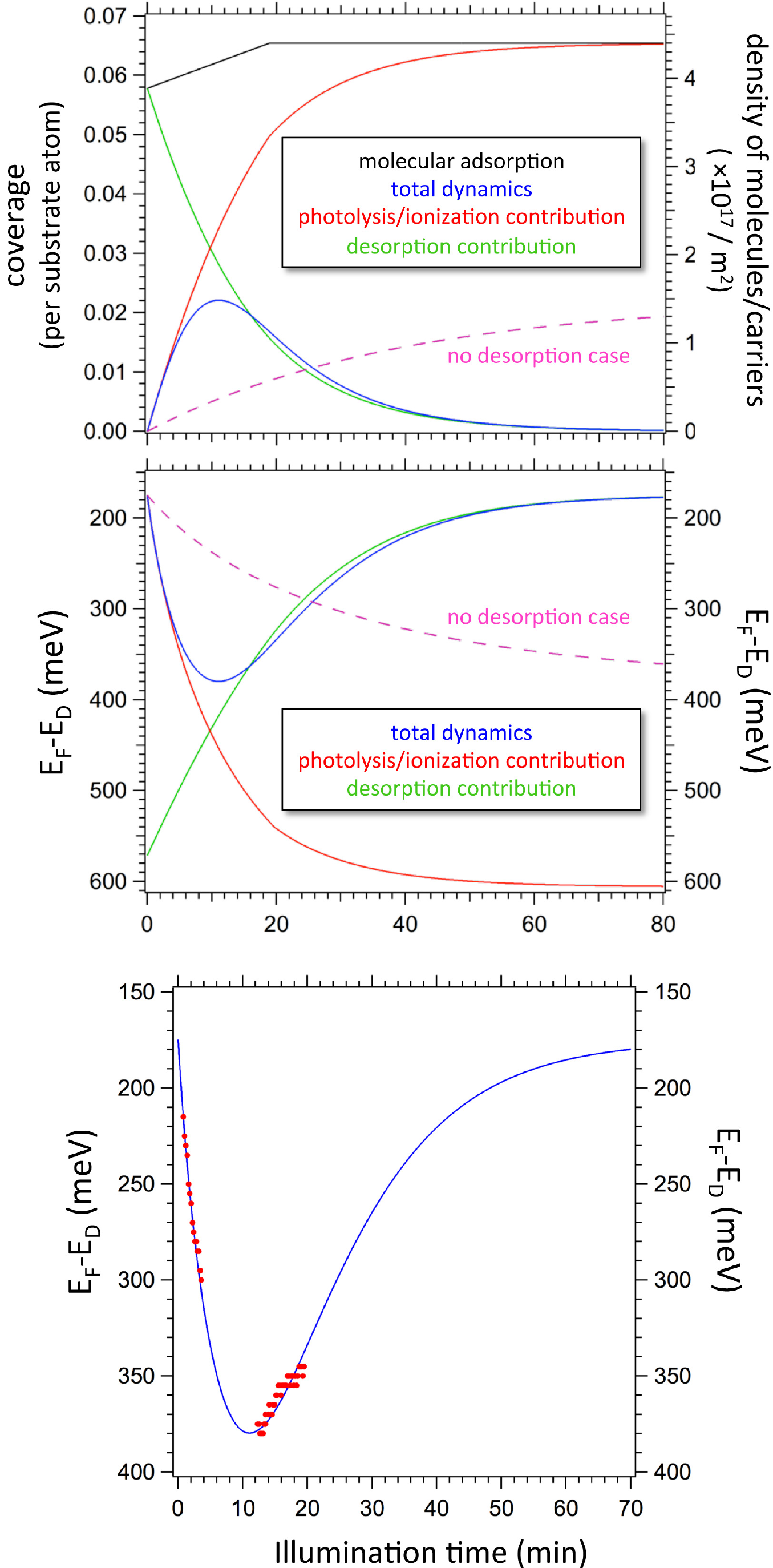}
  \caption{\textbf{A quantitative model of the microscopic procedures behind the surface band bending in 3D TIs.} Top: Time-dependent coverage/density of adsorbed molecules and created charges. Black, red and green curves follow the coverage/density of, respectively, adsorbed molecules, charges created by photo-induced processes and molecules desorbed due to incoming photons. Center: Corresponding time-dependent energy value of the Fermi level with respect to the Dirac point. Red and green curves follow the energy variation due to photo-induced processes of charge creation, and molecular desorption. In both the top and central panels, the blue curve shows the total dynamics under the assumption that photo-stimulated desorption at a local site occurs only after the `activation' of the specific molecule. The dashed curve would describe the total dynamics if in the absence of photo-stimulated desorption the maximum value of surface band bending were to stay unaltered. Bottom: Time-dependent energy variation of the Fermi level in comparison with the experimental data of Fig. 4 showing that the downshift and upshift rates are well reproduced. Details on the model parameters are given in the text.}
\label{figA1}
\end{figure} 

The upper panel of Fig. A1 shows the time-dependent coverage (left axis) and density (right axis) of adsorbed molecules and created charges. The black line corresponds to the coverage/density of adsorbed molecules [equations (15) and (16)]. The red curve follows the coverage and density of charge carriers created by photolysis and photo-ionization [equations (8) and (9)]. In other words, it shows how many of the adsorbed molecules become `activated' by the photon beam. The green curve denotes the decreasing molecular coverage/density due to photo-stimulated desorption [equations (10) and (11)]. Finally, the blue curve shows the total dynamics: the time-dependent coverage/density of the total charge carriers when we consider charge creation and charge removal as photo-induced sequential events at a specific molecular site [equations (12) and (13)]. Best agreement with the experiment is obtained for $t_{\textmd{d}}=8600$ s, a sticking coefficient $s$ of 0.019, a photon flux $F$ of $0.8\times10^{20}$ photons/(s$\cdot$ m$^{2}$), while the values for $N_{\textmd{mol} \, \textmd{max}}$ and $\Theta_{\textmd{mol} \, \textmd{max}}$ have been set to 4$.4\times10^{17}$m$^{\textmd{-}2}$ and $0.065$. These parameter values are in reasonable agreement with the experimental expectations: $t_{\textmd{d}}$ is within the range of 1.5-3.5 hours discussed in the previous subsection, the sticking coefficient is not far from the approximate value of 0.01 calculated from the data of Fig. 4, the photon flux is well below the upper limit of $3.2\times10^{21}$ photons/(s$\cdot$ m$^{2}$) and the values for $N_{\textmd{mol} \, \textmd{max}}$ and $\Theta_{\textmd{mol} \, \textmd{max}}$ are above the lower limits of $1.49\times10^{17}$ m$^{\textmd{-}2}$ and $0.022$, respectively. We note that the curves denoting the density and coverage of adsorbed molecules do not start from zero. A large number of molecules have been already adsorbed on the surface before the beginning of illumination. In these conditions, molecular adsorption reaches the saturation values within 20 mins. The graph shows that the number of adsorbed molecules is the upper limit of the charges created by photo-dissociation/ionization. The last processes determine the total charge and total coverage when $t<5$mins. At later times the number of activated molecules increases, photo-stimulated desorption comes into play, and it finally becomes the dominant process when $t>15$mins. 

\begin{figure}
  \centering
  \includegraphics[width = 8.0 cm]{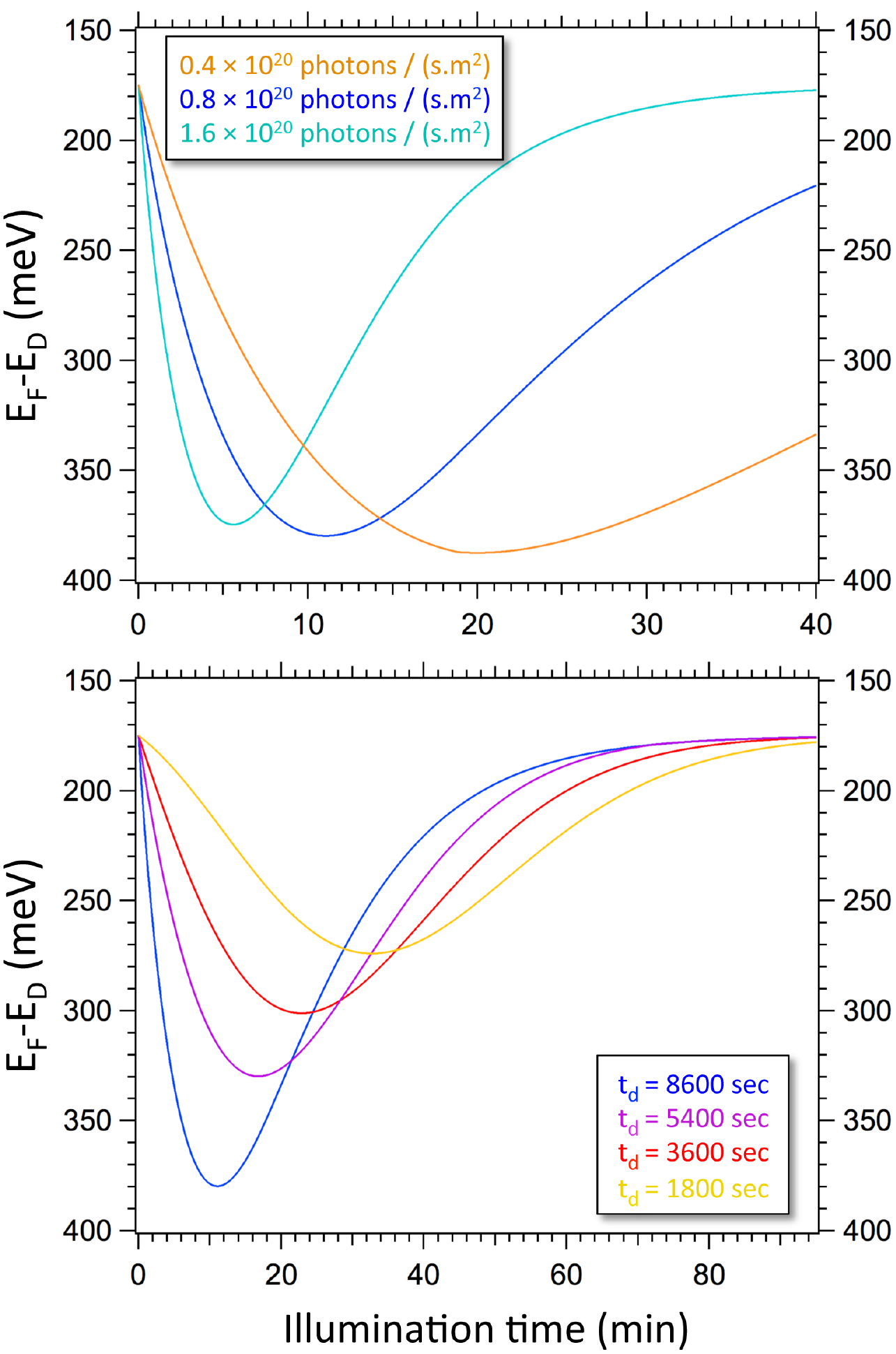}
  \caption{\textbf{The simulated effect of photon flux and surface decoration on the surface band bending in 3D TIs.} Top: Time-dependent energy value of the Fermi level with respect to the Dirac point for various values of the incoming photon flux. Bottom: Time-dependent energy value of the Fermi level with respect to the Dirac point for various time intervals of the surface decoration while the sample was still in the dark. All other parameters of the model are described in the text and they are identical to those used in Fig. A1.}
\label{figA2}
\end{figure} 

The central panel of Fig. A1 follows the complete time-dependence of the simulated Fermi level position with respect to $E_{\textmd{D}}$ [blue line, equation (14)] using the total density of charge carriers that is shown in the top panel. Moreover, the central panel illustrates the time-dependent Fermi level position due to the charge creation processes of photolysis and photo-ionization (green curve) and due to the photo-stimulated desorption (red curve). The red curve has been calculated with $N_{\textmd{charge} \, \textmd{creation}}(t)$ as input for the time-dependent carrier density; in other words, as if a desorption process was not taking place. On the other hand, the green curve has been calculated with $N_{\textmd{mol} \, \textmd{loss}}(t)$ as input for the time-dependent carrier density; in other words, as if each molecule gave rise to a charge carrier before being desorbed and as if photolysis and photo-ionization were not taking place. As expected, the complete time-dependence of the Fermi level position is determined by photolysis/photo-ionization at early times and by photo-stimulated desorption at later times. The dashed curves denote the time-dependent carrier density and coverage to reach the Fermi level position corresponding to maximum band bending in the absence of desorption. In that case, the maximum coverage and carrier density reach the values calculated in the first part of the appendix (i.e. $0.022$ and $1.49\times10^{13}$cm$^{\textmd{-}2}$, respectively). Finally, the bottom panel of Fig. A1 compares the time-dependence of the Fermi level position with respect to $E_{\textmd{D}}$ with the experimental data points of Fig. 4. We note that the rates of both the downshift (data points from location $C$) and the upshift (data points from location $D$) are very well reproduced. 

Fig. A2 shows the predictability of our model. The top panel compares the time-dependence of the Fermi level position for different values of the incoming photon flux. The original value of $F=0.8\times10^{20}$ photons/(s$\cdot$ m$^{2}$) yields the blue curve, while the cyan and orange curves correspond to values of $2F$ and $F/2$, respectively. The time evolution of the energy shifts becomes faster with increasing photon flux as experimentally confirmed in Fig. 4. We note that increasing photon energy could theoretically have the opposite effect from increasing photon flux. This is because the photo-absorption cross-section can strongly decrease when $h\nu$ is higher giving rise to a smaller rate of absorption events [equation (7)]. In practice, the effect of increasing $h\nu$ is more complex. As we have previously discussed in Ref. \onlinecite{Frantzeskakis2015}, increasing $h\nu$ boosts the surface photovoltage process leading to a more effective band flattening through the creation of a larger number of electron-hole pairs.

The bottom panel of Fig. A2 simulates the effect of surface decoration before the first illumination. Blue curve is the original simulation obtained for $t_{\textmd{d}}=8600$ s (2.4 hours). The other curves show the time evolution of the Fermi level when $t_{\textmd{d}}$ changes to 1.5 (purple), 1 (red) and 0.5 (yellow) hours. As time in the dark decreases, the time evolution becomes slower and the total energy shift becomes smaller. These effects are in good agreement with the experimental results presented in Fig. 3. A smaller time in the dark translates into fewer adsorbed molecules at early times and hence into fewer charges from photolysis and photo-ionization before desorption sets in.

The described model of molecular adsorption and photo-stimulated processes can simulate our experimental data in a satisfactory way, and it can moreover correctly predict the system's behavior when certain external parameters are modified. Because of its simplicity, shortcomings are not surprising. As an example, we note the underestimation of the time intervals when the Fermi level position barely changes (e.g. Fig. 4, location $D$). Future improvements require a more refined model that takes into consideration the surface photovoltage process and the possibility of molecular desorption before dissociation or ionization. The inclusion of surface photovoltage in the band flattening process has indeed been the focus of one of our previous studies \cite{Frantzeskakis2015}. Despite the the possibility of more sophisticated simulations, we strongly believe that our simple model already includes the essential physics that govern the time-dependence of the electronic structure of TIs.\\

\section*{Acknowledgements}

This work is part of the research programme Topological Insulators funded by the Foundation for Fundamental Research on Matter (FOM), which is part of the Netherlands Organisation for Scientific Research (NWO).\\

\end{document}